# A Backstepping control strategy for constrained tendon driven robotic finger

Kunal Sanjay Narkhede[1], Aashay Anil Bhise[2], IA Sainul[3], Sankha Deb[4]

[1,2,4]Department of Mechanical Engineering, [3]Advanced Technology Development Centre,
Indian Institute of Technology, Kharagpur.
Kharagpur-721302, India

[1]kunalnarkhede@iitkgp.ac.in, [2]meetaashay@iitkgp.ac.in, [3]sainul@iitkgp.ac.in, [4]sankha.deb@mech.iitkgp.ac.in

*Abstract*—The task of controlling an underactuated robotic finger with a single tendon and a single actuator is difficult. Methods for controlling the class of underactuated systems are available in the literature. However, this particular system does not fall into the class of underactuated system. This paper presents a design change which introduces kinematic constraints into the system, making the system controllable. Backstepping control strategy is used to control the system. Simulation results are presented for single finger driven by a single actuator.

## I. INTRODUCTION

Over the years much effort has been devoted to research and implementation of dexterous robotic grippers. Many designs have been classically proposed for implementation and control of such grippers, but bio-mimicking tendon based, underactuated systems have been constantly discussed throughout the years on various platforms at multiple venues. After careful investigation into the literature, it is found that control of underactuated grippers is very challenging than the fully actuated system. On the other hand, design of underactuated system is much simpler. A mechanical approach to tendon based grippers was given by Ozawa el. al [1], who proposed a compact design method for tendon driven mechanism. A lot of time has been spent in developing mechanical & mathematical models of the systems. The tendon based systems can hence, be broadly classified into three types: fully actuated, under actuated and hybrid actuated Tendon based Mechanisms. Ulrich et al. in [2] proposed a medium – complexity end effector. The control and actuation of this effector was relatively very simple, and it provided reasonable dexterity. The real challenge is not only in designing such an effector, but implementing it to solve real life situations.

In [3], the traditional method of tackling the problem of controlling an underactuated system is explained. The general approach assumes that the coefficient matrix for the control force vector is in a special form which facilitates the division of the equation of the system into component forms namely, unactuated subsystem and actuated subsystem. Then the actuated subsystem is partially linearized into a double integrator and later a diffeomorphism is devised to transform the system co-ordinates and the equations into Z-space. If the underactuated system cannot be exactly feedback linearized, then the system co-ordinates are directly transformed, using a diffeomorphism, into Z-space [4]. The transformed equations are similar to the models after partial feedback linearization but contain complex terms which are diffcult to compute. These complex terms can be considered as uncertainities but cannot be covered by traditional adaptive designs or robust schemes, although a multiple-surface sliding controller can cope with the mismatched uncertainties incorporated with the function approximation techniques to update the uncertainties online[5,6]. Also, the special cascade structure of the system equations, which is prevelant in robots such as [7,8,9], does not allow the use of backstepping method for control expect for sliding or adaptive controller.

Our system, which is a single actuator driving two links of the finger through single tendon, however has a special form of the coefficient matrix for the control force vector. This special form of coefficient matrix of the control force vector does not allow us to seperate the unactuated subsystems and actuated subsystems. We therefore changed the tendon routing of the gripper and remodeled it as a finger subjected to holonomic constraints. The constraints are introduced as per [10] and a backstepping controller is designed for the reduced form of this finger system.

This paper presents a method of controlling the joint variables of a constrained underactuated tendon driven gripper mechanism. The method takes into consideration the motor dynamics [11] of the operating actuator and formulates a backstepping control law [12] for the input voltage. The paper is organized as follows. Section II covers in detail the description of the problem statement and the derives the equations of dynamics of the systems along with stating certain required assumptions. Section III proposes the control scheme and states the stability analysis of the closed loop system. Section IV contains the results which show the performance of the controller while tracking the trajectory. Finally, Section V presents our conclusions.

## II. DYNAMIC EQUATIONS OF KINEMATICALLY CONSTRAINED FINGER

The schematic diagram of the robotic finger is shown in the Fig. 1. The angles swept by the proximal and the distal phalange are $\theta_1$ and $\theta_2$ respectively. Torsional springs are attached to both the joints. The inclusion of torsional springs ensures that the tendon remains taut all the time. The tendon routing is as shown in figure 1. It is assumed that the tendons are inextensible. These tendons are connected to the actuator via a slider. The actuator dynamics taken into consideration is of a DC Motor. If the above assumption holds true then this particular tendon routing and slider arrangement constrain the kinematic motion of the finger. The constraint motion can be realized by assuming that the displacements of the slider and the tendons connecting to both the phalanges are the same. The design of the slider and tendon routing is inspired from [13].

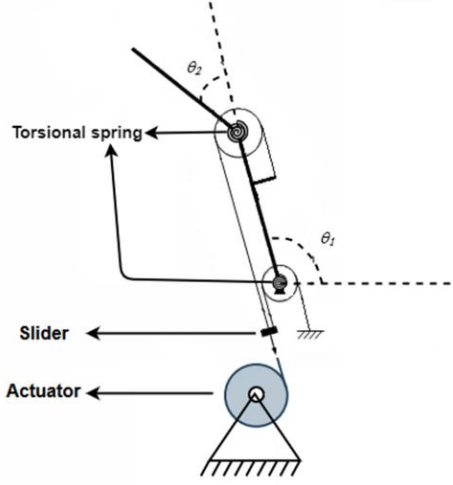

Figure 1. The schematic diagram of robotic finger

## A. Dynamic Model of finger

First, let us consider the dynamic equations of **2** degrees of freedom planer finger subjected to **1** holonomic kinematic constraint. These equations are based on the Euler-Lagrangian formulation and can be represented as:

$$M(\theta)\ddot{\theta}+C(\theta,\dot{\theta})\dot{\theta}+G(\theta)+K\theta=\tau+A^T(\theta)\lambda \quad (1)$$

where $\theta \in R^2$ is the angular position vector. $\tau \in R^2$ is the generalized torque vector, $M(\theta) \in R^{2*2}$ is symmetric and positive definite inertia matrix, $C(\theta,\dot{\theta}) \in R^{2*2}$ is the the centripetal and coriolis torque matrix, $G(\theta) \in R^2$ is the gravitational torque vector, $K \in R^{2*2}$ is constant stiffness matrix which is diagonal, $A^T(\theta)\lambda$ is the constraint force term and $\lambda$ lagrange multiplier. For details, refer to [14]. The generalized torque vector can be expressed in terms of tendon forces and pulley radii as follows:

$$\tau=\begin{bmatrix}r_1*f_1\\r_2*f_2\end{bmatrix} \quad (2)$$

where $r_1$ and $r_2$ are the pulley radii and $f_1$ and $f_2$ are the tendon forces in the proximal and distal phalanxes respectively.

## B. Model Reduction

The finger is subjected to kinematic constraints defined by

$$\theta_1-\left(\frac{r_2}{r_1}\right)*\theta_2=0 \quad (3)$$

where $\theta_1$ and $\theta_2$ are the joint angles. Differentiating the above equations with time we get

$$A(\theta)\dot{\theta}=0 \quad (4)$$

where $A(\theta)=\begin{bmatrix}1 & -\frac{r_2}{r_1}\end{bmatrix}$ and $\theta=\begin{bmatrix}\theta_1\\\theta_2\end{bmatrix}$. From (3) we can write

$$\theta=\begin{bmatrix}1\\\frac{r_1}{r_2}\end{bmatrix}\theta_1=D\theta_1 \quad (5)$$

As **D** lies in the null space of **A**, the following equation is satisfied

$$AD=D^TA^T=0 \quad (6)$$

Further differentiating (5) we get

$$\ddot{\theta}=D\ddot{\theta}_1 \quad (7)$$

From (1) and (7), the reduced dynamic equation of the system can be written as

$$M'(\theta_1)\ddot{\theta}_1+C'(\theta_1,\dot{\theta}_1)\dot{\theta}_1+G'(\theta_1)+K'(\theta_1)=r_1(f_1+f_2) \quad (8)$$

where,

$$M'=D^TMD, \ C'=D^TCD.$$

$$G'=D^TG, K'=D^TK.$$

It should be noted that the above equation is a scalar equation. The dynamics of slider and actuator is given by the equations:

$$m\ddot{x}=f-(f_1+f_2) \quad (9)$$

$$J\ddot{\phi}+B\dot{\phi}+r_af=\tau_a \quad (10)$$

where $\phi$ is the angular position of the actuator, $\tau_a$ is the actuator torque, **J** is the actuator inertia, **B** is the actuator damping, $r_a$ is the radius of pulley mounted on the actuator. The slider displacement, the actuator angular displacement $\phi$ and the joint angles are related as:

$$x=r_a\phi=r_1\theta_1=r_2\theta_2 \quad (11)$$

Substituting the above relation in (9) and eliminating **f** we get

$$(J+mr_a^2)\ddot{\phi}+B\dot{\phi}+r_a(f_1+f_2)=\tau_a \quad (12)$$

Now eliminating $f_1$ and $f_2$ from equation (8) and (9) and substituting $\phi=(r_1/r_a)\theta_1$ we get:

$$((r_a/r_1)M'(\theta_1) + J + mr_a^2)\ddot{\theta}_1+((r_a/r_1)C'(\theta_1,\dot{\theta}_1) + B)\dot{\theta}_1$$
$$+(r_a/r_1)G'(\theta_1)+(r_a/r_1)K'(\theta_1)=\tau_a \quad (13)$$

$$M''(\theta_1)\ddot{\theta}_1+C''(\theta_1,\dot{\theta}_1)+G''(\theta_1)+K''(\theta_1)=\tau_a \quad (14)$$

Now let us consider the actuator electrical circuit. The relation between the actuator angle and $\phi$, the torque generated and the current flowing through the circuit is given by:

$$L*\frac{di}{dt}+R_a i+K_b \dot{\phi}=E \quad (15)$$

$$\tau_a = K_t I \quad (16)$$

where $i$ is the current flowing through the circuit. $R_a$ is the armature resistance, $K_b$ is the back-emf constant, $L$ is the inductance, $E$ is the input voltage signal and $K_t$ is the actuator torque constant. From (15) and (16) and substituting $\phi=(r_1/r_a)\theta_1$ we get:

$$\frac{L}{K_t}*\frac{d\tau_a}{dt}+\frac{R_a}{K_t}\tau_a+\frac{K_b r_1}{r_a}\dot{\theta}_1=E \quad (17)$$

### III. BACKSTEPPING CONTROL STRATEGY

In this section we have used the backstepping control law to track the desired trajectory. We have considered voltage signal given to the actuator as input and the proximal joint angle as the output. Lyapunov stability analysis is used to derive the control law for the system. Let $x_1=\theta_1$, $x_2=\dot{\theta}_1$ and $x_3=\tau_a$. Expressing (14) and (17) in state space we get:

$$\dot{x}_1 = x_2$$
$$\dot{x}_2 = f(x_1,x_2)+g(x_1,x_2)x_3$$
$$\dot{x}_3 = -\frac{R_a}{L}x_3 - \frac{K_t K_b r_1}{r_a L}x_2 + \frac{K_t}{L}E \quad (18)$$

Where $f(x_1,x_2)=\frac{-C''(x_1,x_2)-G''(\theta_1)-K''(\theta_1)}{M''(x_1)}$ & $g(x_1,x_2)=\frac{1}{M''(x_1)}$.
To express the above equation in standard form required to implement backstepping control, we design our control law as follows:

$$E=\frac{L}{K_t}*\left(\frac{R_a}{L}x_3+\frac{K_t K_b r_1}{r_a L}x_2+u\right) \quad (19)$$

So now the equation becomes:

$$\dot{x}_1 = x_2$$
$$\dot{x}_2 = f(x_1,x_2)+g(x_1,x_2)x_3$$
$$\dot{x}_3 = u \quad (20)$$

Now, we define the error signal in $x_1$ as $e=x_1-x_{1d}$ and an error surface as $s=\dot{e}+\lambda e$, where $\lambda$ is a positive constant. Differentiating the error surface, we get:

$$\dot{s}=\ddot{e}+\lambda \dot{e}$$
$$\dot{s}=\ddot{x}_1 - \ddot{x}_{1d}+\lambda \dot{e}$$
$$\dot{s}=f(x_1,x_2)+g(x_1,x_2)x_3 - \ddot{x}_{1d}+\lambda \dot{e} \quad (21)$$

Here we define a virtual control $x_{3d}$ and define the error in virtual control as $\eta=x_3-x_{3d}$. Equation (21) can be written in terms of the virtual control as:

$$\dot{s}=f(x_1,x_2)+g(x_1,x_2)\eta+g(x_1,x_2)x_{3d} - \ddot{x}_{1d}+\lambda \dot{e} \quad (22)$$

Now in order to make $s$ zero at steady state, we select the virtual control as:

$$x_{3d}=\frac{1}{g(x_1,x_2)}(-f(x_1,x_2)+\ddot{x}_{1d}-\lambda \dot{e}-k_1 s) \quad (23)$$

So after applying this control, the close loop error dynamics is given by:

$$\dot{s} = g(x_1, x_2)\eta - k_1 s \quad (24)$$

where $k_1$ is a positive constant. Differentiating the error in virtual control we get:

$$\dot{\eta}=\dot{x}_3-\dot{x}_{3d}$$
$$\dot{\eta}=u-\dot{x}_{3d} \quad (25)$$

So we select the control input $u$ as follows:

$$u = \dot{x}_{3d}-k_2\eta-g(x_1,x_2)s \quad (26)$$

So the close loop dynamics of $\eta$ is given by

$$\dot{\eta}=-k_2\eta-g(x_1,x_2)s \quad (27)$$

where, $k_2$ is a positive constant and $-g(x_1,x_2)s$ is derived using Lyapunov stability analysis. The stability analysis of the control law is done using Lyapunov stability theory. We define a positive definite Lyapunov function as follows:

$$V=\frac{1}{2}(s^2+\eta^2) \quad (28)$$

Differentiating the above equation, we get,

$$\dot{V}=s\dot{s}+\eta\dot{\eta} \quad (29)$$

Substituting the values of $\dot{s}$ and $\dot{\eta}$ back into the equation we get

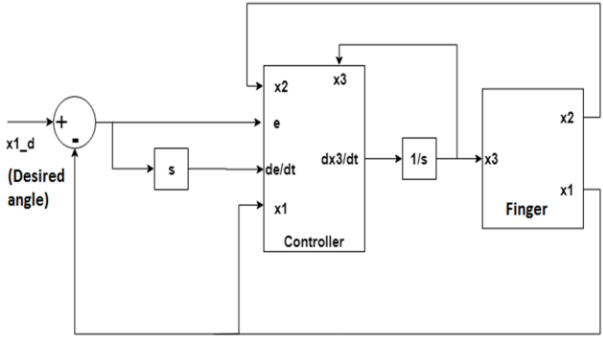

Figure 2. Block diagram of the control system

$$\dot{V} = s(g(x_1,x_2)\eta - k_1 s) + \eta(-k_2\eta - g(x_1,x_2)s)$$

$$\dot{V} = -k_1 s^2 - k_2 \eta^2 \quad (30)$$

The above equation is negative definite. This proves the stability of the proposed control law. The control law can be stated as follows:

$$E = \frac{L}{K_t} * \left( \frac{R_a}{L} x_3 + \frac{K_t K_b r_1}{r_a L} x_2 + \dot{x}_{3d} - k_2 \eta - g(x_1,x_2)s \right) \quad (31)$$

## IV. SIMULATION RESULTS

Fig. 2. shows the block diagram of the control system on which the simulations we performed. We have performed numerical simulation in MATLAB Simulink to test the performance of the proposed control law. The system parameters used in simulation are as follows: $m_1 = 50g$, $m_2 = 40g$, $l_1 = 60mm$, $l_2 = 40mm$. The parameters of the actuator are as follows: $J = 1.5*10^{-4} Kg\text{-}m^2$, $B = 0.03 Nms/rad$. The slider mass is $m = 20g$ and the actuator pulley radius is $r_a = 10mm$. The parameters of the controller are: $\lambda = 3.4$, $k_1 = 28$ and $k_2 = 40$. The step size for simulation is 0.01s.

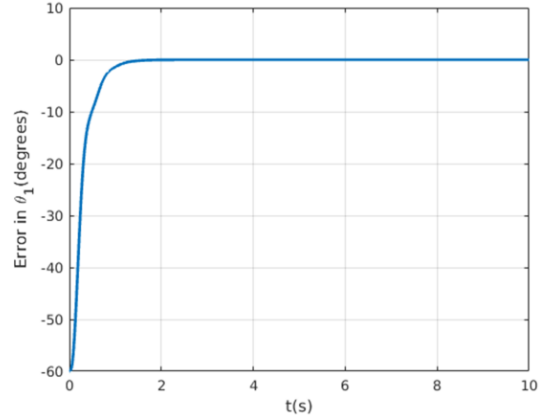

Figure 4. Error in proximal link angle

### A. Step Response

Fig. 3 shows the step response of the system. A step input with a value of 60° is provided and the resulting response is recorded. As observed from the figure, the response reaches its desired value within 1.5s. Also there is no overshoot. This shows that the control system is critically damped. It can also be observed from Fig. 4 that the error in proximal joint angle tends to zero within 1.5 seconds.

The error in virtual control settles faster as compared to error in proximal joint angle as observed in Fig. 5. This is evident from the fact that the value of $k_2$ is more than $k_1$. Also the close loop error dynamic equations show that in order to make the error in proximal joint angle zero, the error in virtual control should first become zero. This proves that the controller behaves exactly as designed.

### B. Cubic Polynomial Trajectory Tracking

Here we have given cubic polynomial trajectory as input to the system. The equation describing the polynomial in time is given by: $\theta_1(t) = -0.0021t^3 + 0.0314t^2$. Fig. 6 shows the cubic polynomial trajectory tracking response of the controller. As observed from the figure the error in the response is negligible. Fig. 7 shows the variation of error in joint angle with time. The order of magnitude is about 0.01 degrees which is very low.

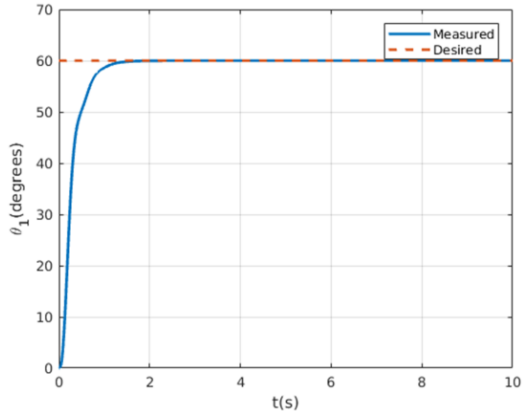

Figure 3. Step Response of proximal link angle

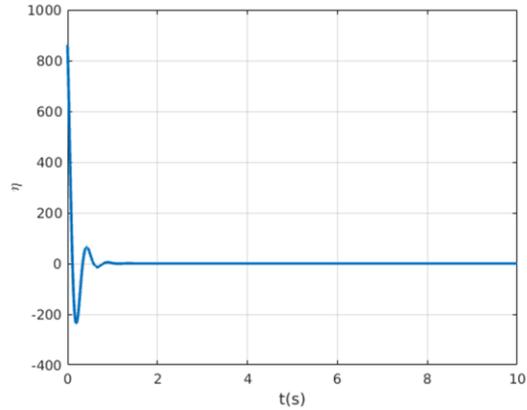

Figure 5. Error in virtual control

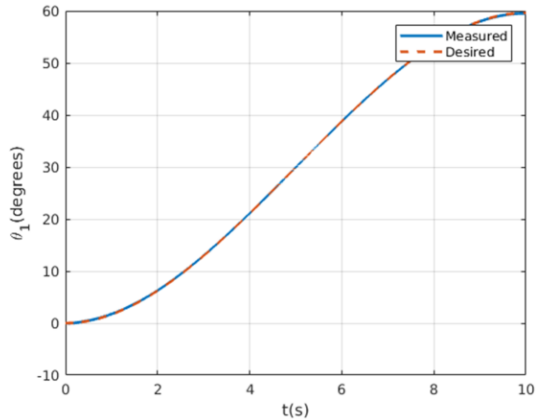

Figure 6.  Cubic polyniomial trajectory tracking

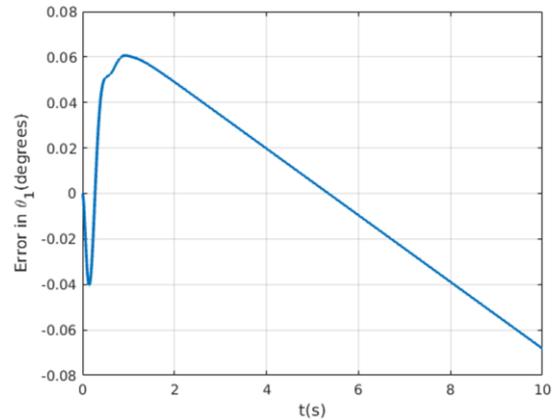

Figure 7.  Error in cubic polyniomial trajectory

## V. CONCLUSION

We have implemented a backstepping control strategy on the robotic finger with kinematic constraints. Actuator dynamics has been included in the dynamic model so that the theoretical formulations can be implemented on a real system. The stability of the controller has been proved mathematically using Lyapunov stability theory. While simulating the controller, the system parameter values are realistic and the performance of the controller shows the possibility of practical implementation. Implementation of this approach on a practical model is a part of the ongoing research. In future, emphasis will be given on adding multiple such grippers to a manipulator and focusing on problems pertaining to grasping of objects.